\documentclass[a4paper]{jpconf}
\usepackage{graphicx}
\usepackage{cite}
\usepackage{gensymb}
\usepackage{amsmath}
\usepackage{amssymb}

\begin{document}
\title{Satellite monitoring of atmospheric temperature profiles and cloud cover over the Yakutsk EAS array and the TAIGA observatory}

\author{Anatoly A. Lagutin, Nikolay V. Volkov, Andrey P. Zhukov, Konstantin M. Makushev, Alexander A. Maslov, Egor Yu. Mordvin, Roman I. Raikin, Tatyana L. Serebryakova, Vladimir V. Sinitsin}

\address{Altai State University, 61 Lenin St., Barnaul, Russian Federation, 656049}

\ead{lagutin@theory.asu.ru}

\begin{abstract}
We consider the problem of night-time atmosphere monitoring at the locations of the Yakutsk EAS array and the TAIGA observatory for the annual October--April period of observations at these facilities. It is shown that the use of data from instruments aboard Terra, Aqua, Suomi-NPP and NOAA-20 satellites,  received by the ground stations of the Altai State University, Russia, and processed to Level 2 (retrieval of geophysical parameters of the atmosphere), gives the possibility of the near-real-time  monitoring observations of the cloud cover structure and the temperature profiles at night with a spatial resolution and a frequency sufficient for meteorological correction of the detector readings.

\end{abstract}

\section{Introduction}

Today, extensive air showers (EAS) are the only source of information about physical characteristics of the  very high energy primary cosmic radiation ($E>10^{15}$ eV).
Composition, energy spectrum and arrival direction of a primary particle can be retrieved on the basis of fluorescence/Cherenkov light and lateral distribution of secondary particles which are measured by EAS array detectors spread on large areas. 

However, such a retrieving is interfered by variations of properties of the atmosphere where cascade processes develop. Experimental data for different observed showers correspond to different states of the atmosphere. For example, the lateral distribution of electrons generated by air shower is affected by the air density distribution $\rho(t)$ throughout the atmospheric depth $t$, or by temperature distribution due to  relationship between density and temperature $\rho(t) \sim 1/T(t)$. Thus variations of the temperature profile change the lateral distribution of particle density. Regarding fluorescence/Cherenkov detectors readings, another essential factor is the presence of cloud structures in the atmosphere over the array area.

The main goal of this work is to study a capability to measure variations of temperature profiles of the atmosphere and presence of night-time cloud cover over the locations of the Yakutsk EAS array~\cite{ Ivanov:2013} and the TAIGA observatory~\cite{Kuzmichev:2018} in the near-real-time using data from satellite instruments. 

\section{Satellite observations}

Temperature profiles of the atmosphere required for the meteorological correction of experimental data at the locations of the Yakutsk EAS array (61.7\degree N, 129.4\degree E) and the TAIGA observatory (51.8\degree N, 103.1\degree E) have been obtained using data from the hyperspectral infrared radiometer AIRS (Atmospheric Infrared Sounder)~\cite{Aumann:2003} aboard the Aqua satellite~\cite{Parkinson:2003}. AIRS was specially designed to solve the key problem of satellite meteorology --- measurement of the Earth's atmospheric temperature profiles with retrieval accuracy of better than 1 K in the lower troposphere under clear and partly cloudy conditions on a global scale. 

AIRS software packages v.5.0.22.100 (Level 1B) and v.6.0.12.101 (Level 2) have been used to retrieve the characteristics of the atmosphere and the underlying surface of the Earth. The input information for these packages is data in the PDS (Production Data Set) format. These PDS files are created in the Remote Sensing Center  of the Altai State University after unpacking the full raw data of Aqua satellite, received by EOScan and UniScan-24 ground stations in broadcast mode followed by using the RT-STPS (Real-time Software Telemetry Processing System) package.

Night-time cloud cover monitoring over the EAS arrays mentioned above was carried out using MOD35/MYD35 and VIIRS Cloud Mask products  obtained in the Remote Sensing Center from the data of the MODIS/Terra-Aqua~\cite{Salomonson:1989} and VIIRS/SNPP-NOAA-20~\cite{Hillger:2013,Goldberg:2013} instruments, respectively.
Note that MODIS channels used for cloud mask retrieval have nadir spatial resolution of 1000 m, while VIIRS channels have resolution of 750 m. 
 
\section{Atmospheric temperature profiles from AIRS/Aqua data}

As a result of daily satellite measurements, we have created a database of temperature profiles over the above mentioned EAS experiments for the period of October, 2016--April, 2017.

\begin{figure}[bh]
\includegraphics[width=\textwidth]{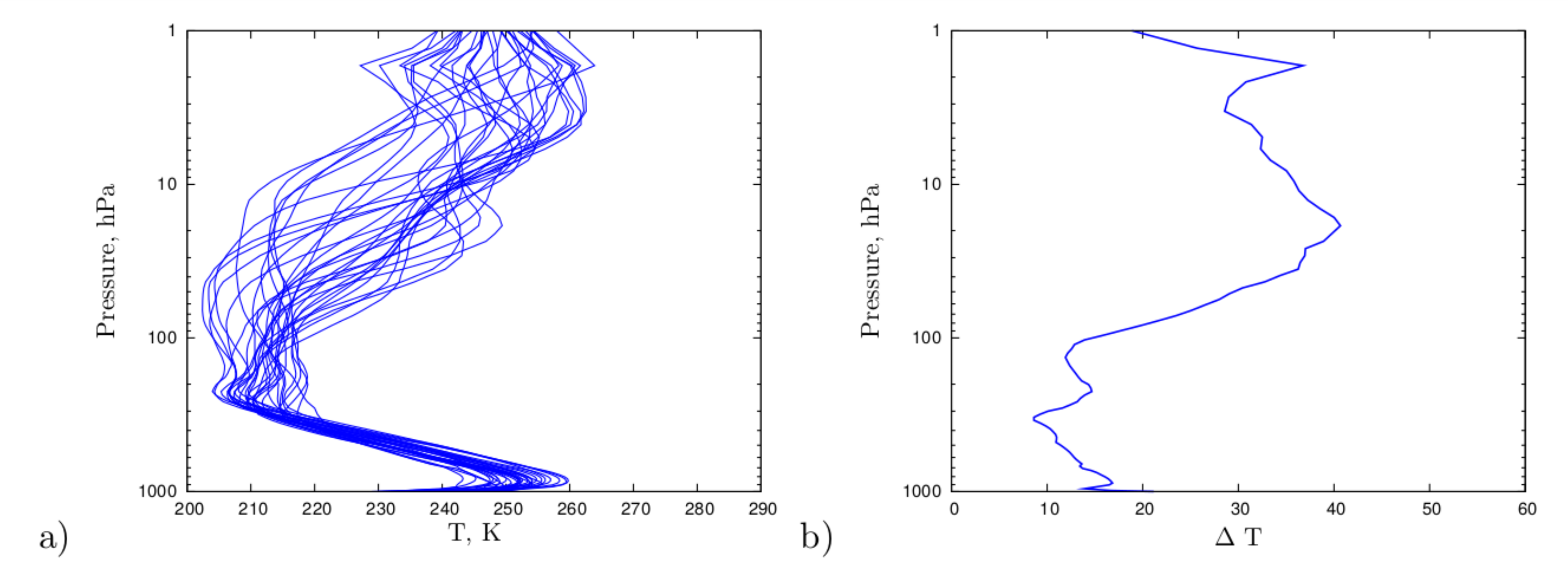}
\caption{Daily temperature profiles (a) and the range of temperature variations (b) at night-time over the Yakutsk EAS array for January, 2017}
\label{temp_yakutsk}
\end{figure}

Figure~\ref{temp_yakutsk}a) shows the results of measurements of temperature profiles for night swaths of AIRS/Aqua at the location of Yakutsk EAS array in January, 2017. The range of temperature variations in the atmosphere in this zone is shown in Figure~\ref{temp_yakutsk}b). Similar data were obtained for the TAIGA observatory. 

\begin{figure}[t]   
\includegraphics[width=\textwidth]{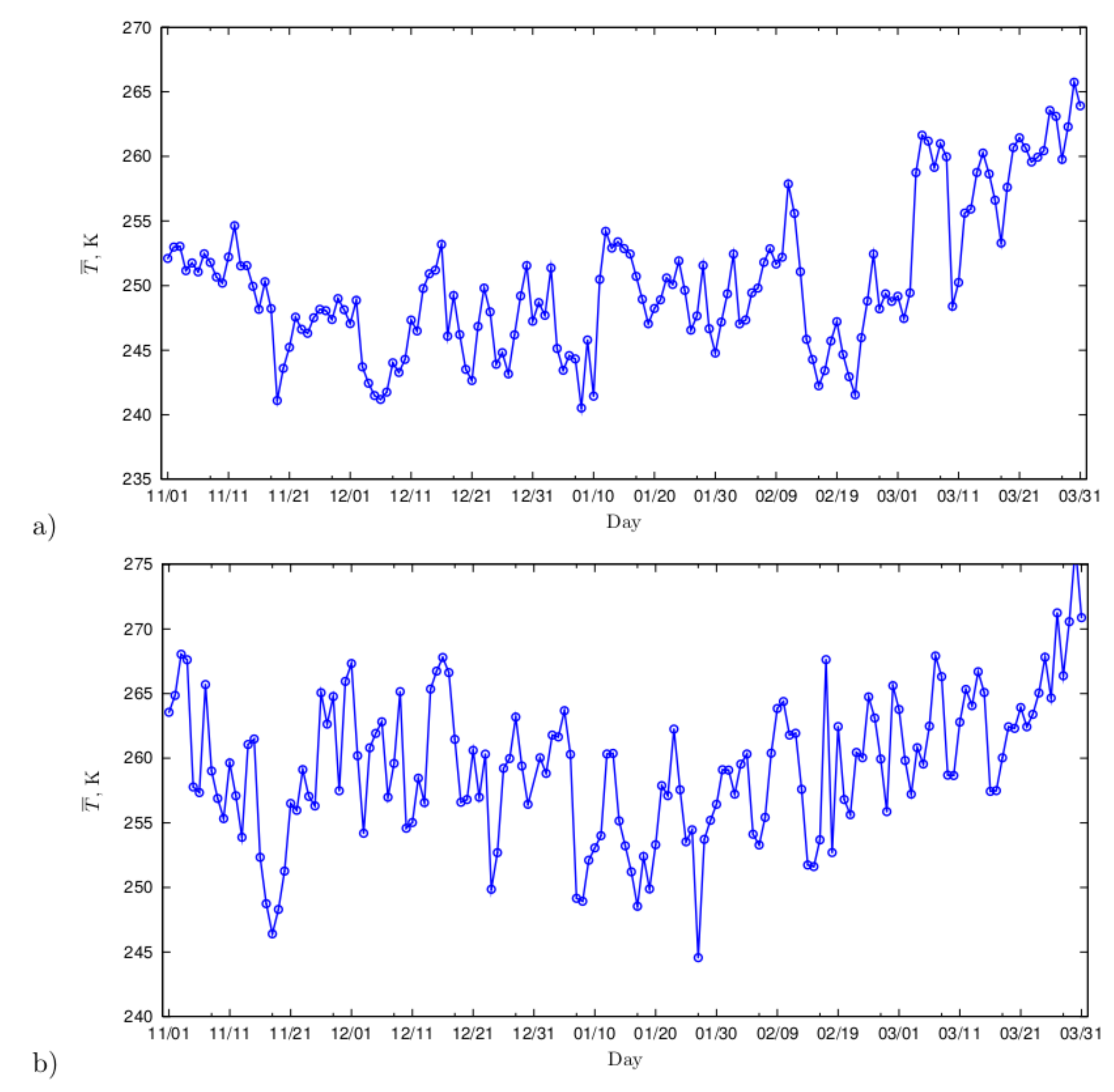}
      \caption{Variations of the average temperatures in a layer of 300 hPa above the observation level from November, 2016 to March, 2017 (a) at the location of the Yakutsk EAS array, b) at the location of the TAIGA observatory}
\label{vartemp}
\end{figure}

Variations of the average temperature values in a layer of 300 hPa above the observation level from November, 2016 to March, 2017 are shown in Figure~\ref{vartemp}. AIRS/Aqua data show that such variations in the lower troposphere can reach up to 30 degrees.

\begin{figure}[!hbt]    
\includegraphics[width=\textwidth]{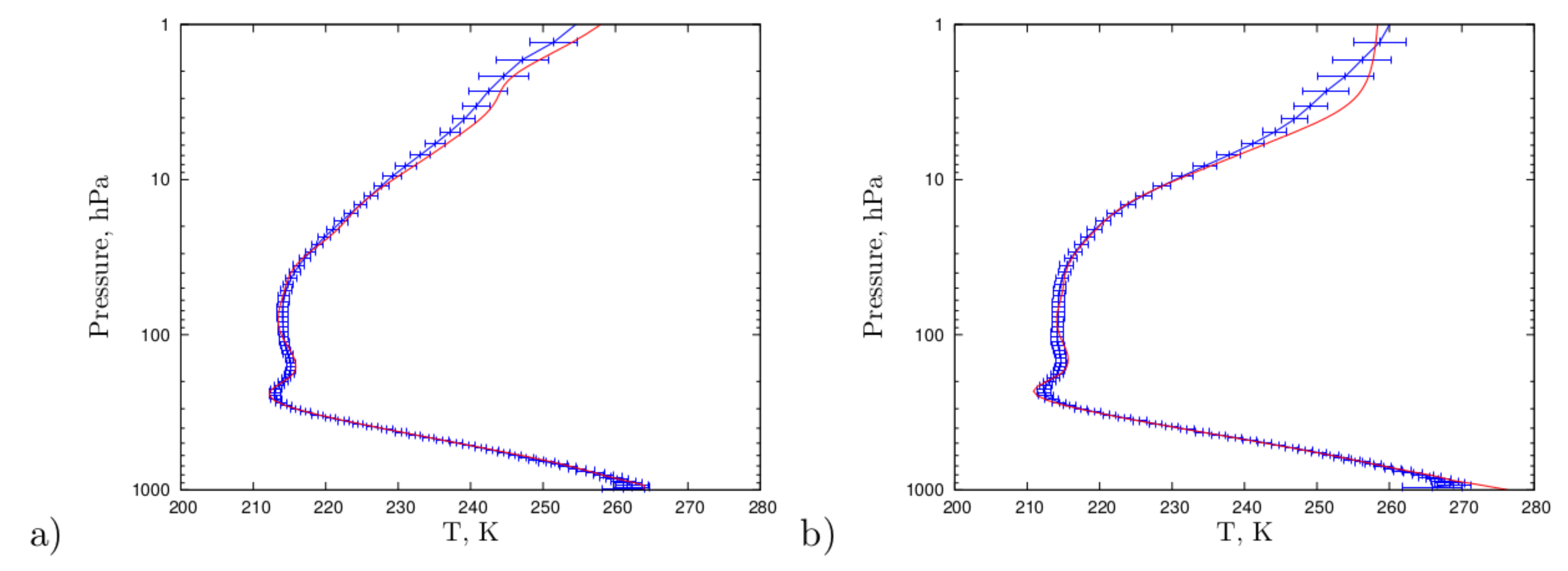}
      \caption{Average temperature profile of the atmosphere in March, 2017 (a) at the location of the Yakutsk EAS array, b) at the location of the TAIGA observatory. The AIRS/Aqua data are plotted by the blue line; the results of the ERA-Interim reanalysis are plotted by red line. Intervals show retrieval errors of the monthly average temperature values according to the AIRS/Aqua}
\label{temp_compare}
\end{figure}

We have made additional comparisons of the satellite data with results of the ERA-Interim reanalysis~\cite{Dee:2011}. The data presented in Figure~\ref{temp_compare} show that satellite observations are in good agreement with the results of the reanalysis.

\section{Cloudiness according to MODIS/Terra-Aqua and VIIRS/SNPP-NOAA-20 data}

\begin{figure}[!hbt]    
\includegraphics[width=\textwidth]{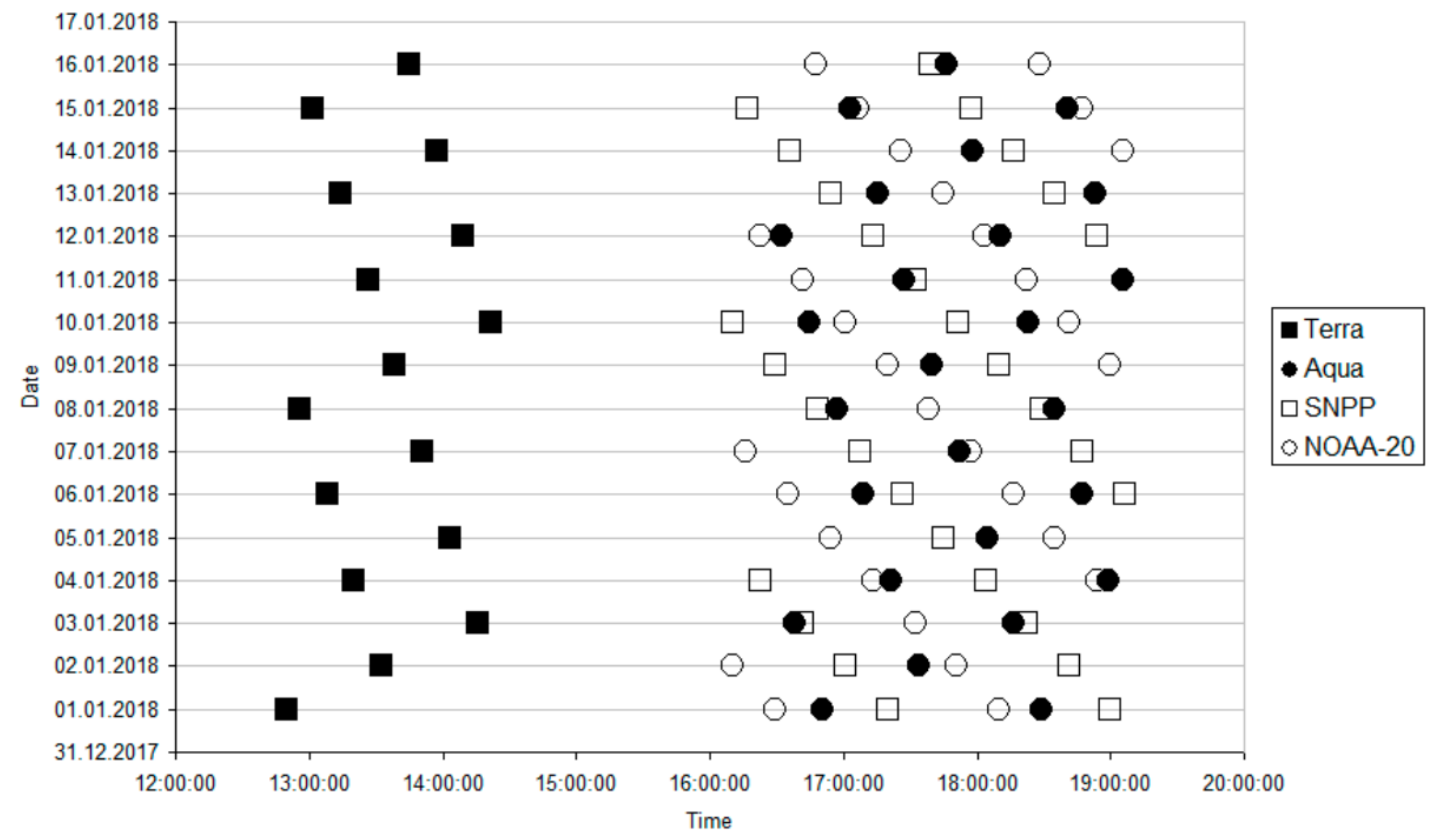}
      \caption{The time when the Yakutsk EAS array is covered by radiometers MODIS/Terra-Aqua and VIIRS/SNPP-NOAA-20 in the 16-day cycle of MODIS in the period of 13:00-19:00 UTC (22:00-04:00 local time)}
\label{covertime_yakutsk}
\end{figure}

Figure~\ref{covertime_yakutsk} shows the time when the area of Yakutsk EAS  array 
is observed by radiometers MODIS/Terra-Aqua and VIIRS/SNPP-NOAA-20 in the 16-day cycle of MODIS for the period 01.01.2018--16.01.2018. The analysis of that and others 16-day MODIS cycles  showed that presently satellite instruments provide a sufficient frequency of cloud cover observations over the areas of the Yakutsk EAS array and the TAIGA observatory during the periods 13:00-14:00 UTC and 16:00-20:00 UTC. The period of 14:00-16:00 UTC should be controlled by surface instruments.

\begin{figure}[!hbt]    
\includegraphics[width=\textwidth]{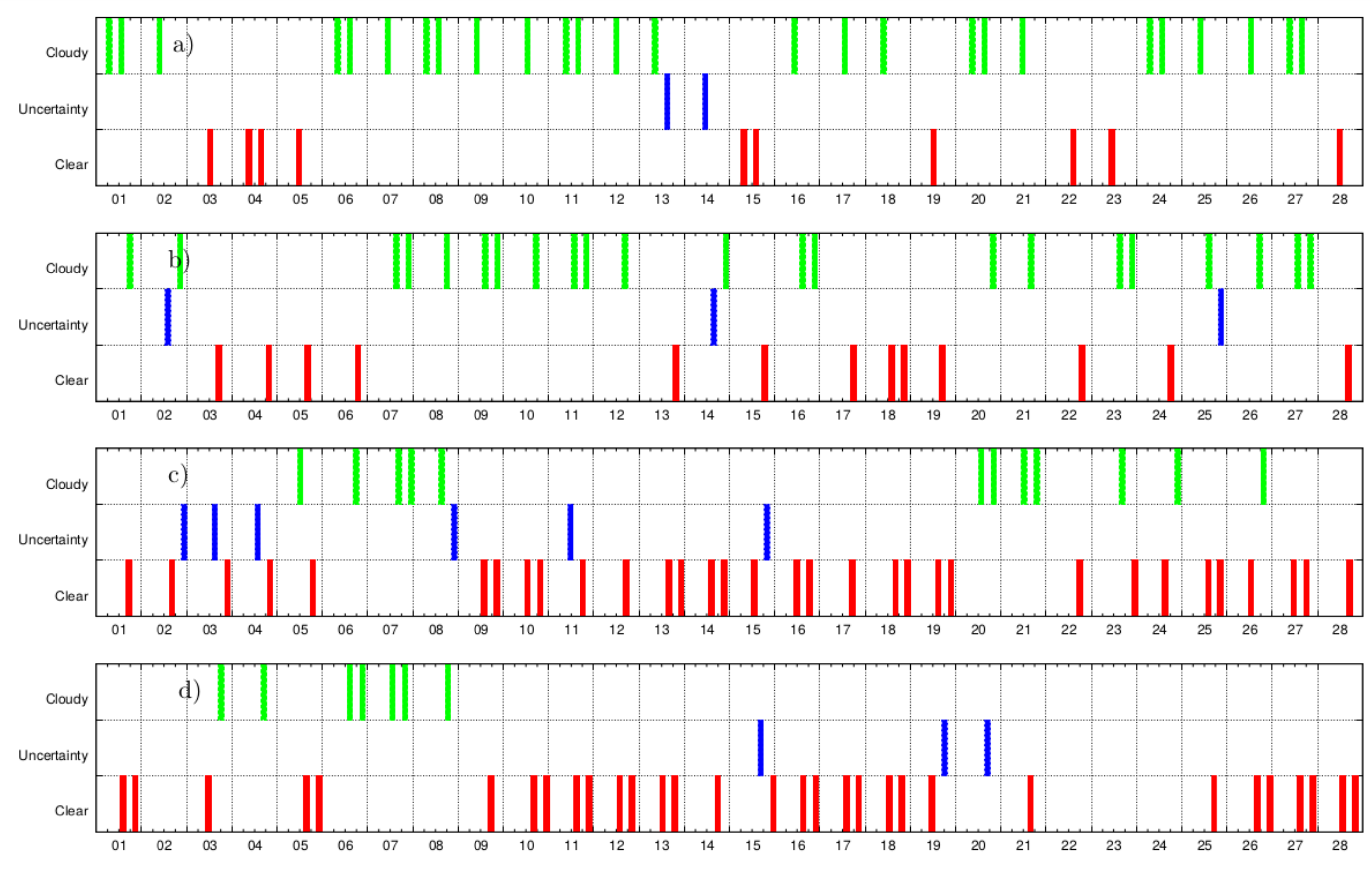}
      \caption{Cloud cover over 2 km by 2 km area over TAIGA observatory in February 2018 according to data of a) MODIS/Terra; b) MODIS/Aqua; c) VIIRS/SNPP; d) VIIRS/NOAA-20}
\label{clouds_taiga}
\end{figure}

Figure~\ref{clouds_taiga} shows the behavior of average cloud mask over 2 km by 2 km area covering TAIGA observatory in February 2018. It is easy to see that according to data of MODIS/Terra-Aqua about a half of measurements indicates cloudiness, while according to data of VIIRS/SNPP-NOAA-20 the most part of the area is free of clouds.  

Note that although the definition of a cloud mask is one of four possible levels of confidence~--- confidently clear, possibly clear, uncertain (probably cloudy) and confidently cloudy, for the convenience of time-series analysis, in this paper the first two of them were combined into one~--- clear.

\begin{figure}[!hbt]    
\includegraphics[width=\textwidth]{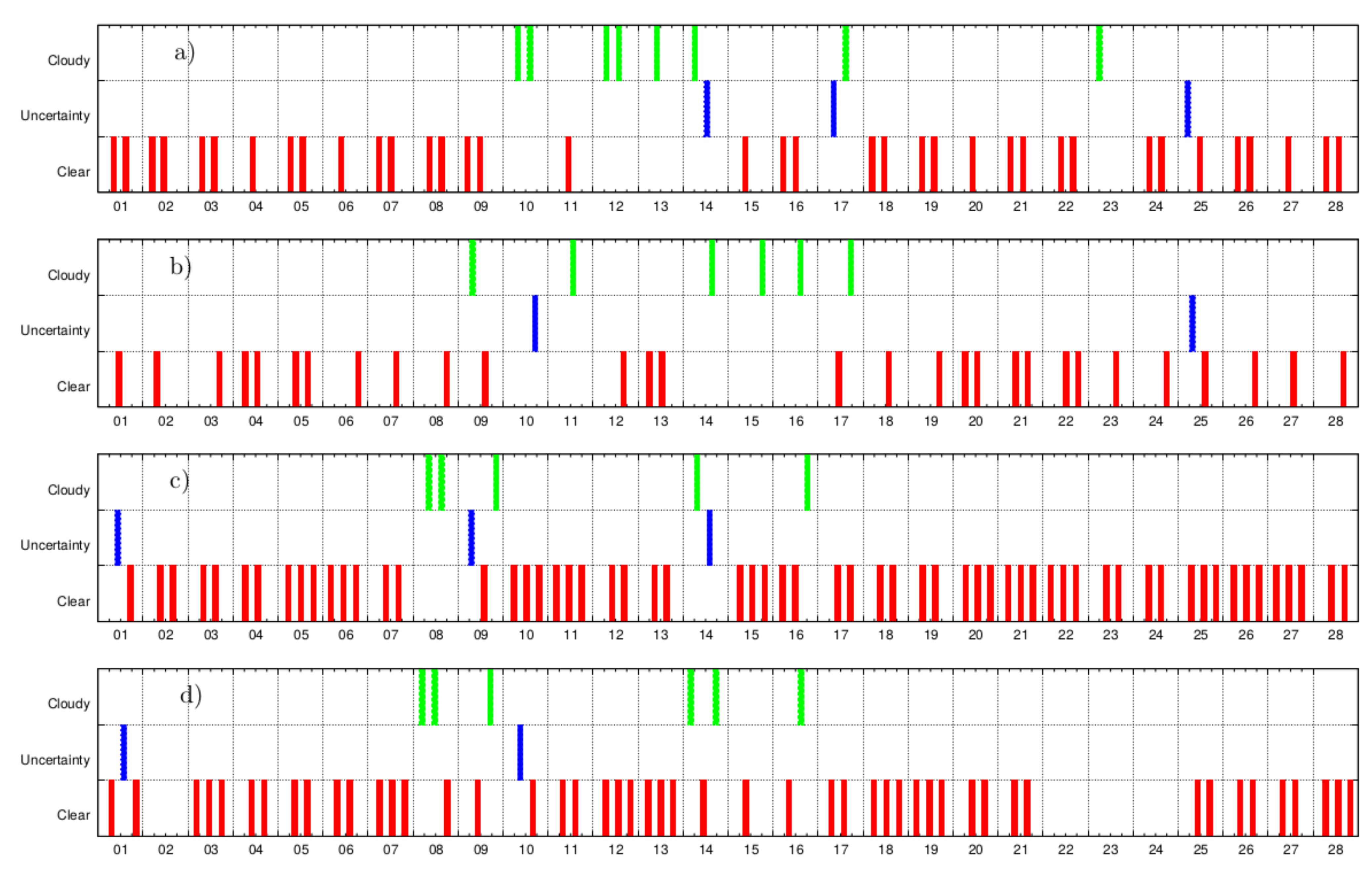}
      \caption{Cloud cover over Yakutsk EAS array area 3 km by 4 km in February 2018 according to data of a) MODIS/Terra; b) MODIS/Aqua c) VIIRS/SNPP; d) VIIRS/NOAA-20}
\label{clouds_yakutsk}
\end{figure}

The behavior of average cloud mask over 3 km by 4 km area covering Yakutsk EAS array in February 2018 is shown on Figure~\ref{clouds_yakutsk}. For this area MODIS/Terra-Aqua and VIIRS/SNPP-NOAA-20 results comply with each other much better. According to satellite observations, this area was free of clouds during the most part of the period.

\section{Conclusions}
Using data of hyperspectrometer AIRS/Aqua received by the Remote Sensing Center of Altai State University several times a day, the database of daily temperature profiles at the locations of the Yakutsk EAS array and the TAIGA observatory have been created for the period October, 2016--April, 2017. It was found that in the layer of 300 hPa above the observation level for the period from November 1st, 2016 to March 31st, 2017 the variations of the average atmospheric temperature were equal to 25 K and 30 K, respectively.

A problem of monitoring of cloud cover, which affects an appropriate detection of Cherenkov radiation, in the areas of Yakutsk EAS array and TAIGA observatory at night-time during annual observations from October to April has been discussed. The main sources of quantitative information on the state of the atmosphere were data from MODIS/Terra-Aqua and VIIRS/SNPP-NOAA-20 satellite instruments. 

It is shown, that utilization of data from instruments aboard Terra, Aqua, SNPP and NOAA-20 satellites, received by ground stations of Altai State University and processed up to level 2 (retrieval of geophysical parameters of atmosphere), gives opportunity to carry out monitoring observations of cloud cover structure at night-time with high spatial resolution and frequency which is sufficient to make cloud correction of detector arrays readings.

Analysis of temporal sampling of satellite monitoring shows that the area of TAIGA observatory is covered by satellite observations 5-7 times during the night, while area of Yakutsk EAS array is observed by satellites 8-10 times per night.  

Intercomparison of cloud mask products from MODIS/Terra-Aqua and VIIRS/SNPP-NOAA-20 algorithms leads to their significant difference for TAIGA observatory. However, for the area over Yakutsk EAS array the results correspond much better. In our view, an additional verification of these products in the areas of TAIGA observatory could be made by their comparison with VIIRS  Day/Night Band  data in full moon period.

\ack
This work was supported in part by the Ministry of Science and Higher Education of the Russian Federation (state assignment for the fundamental and applied research performed at Altai State University, project No 3.5939.2017/8.9)

\end{document}